\documentclass[twocolumn, showpacs,prl]{revtex4}
\usepackage{graphicx}
\usepackage{dcolumn}
\usepackage{bm}

\begin{document}

\title{Effective mass suppression in dilute, spin-polarized two-dimensional electron systems}

\date{\today}

\author{Medini\ Padmanabhan}

\author{T.\ Gokmen}

\author{N.C.\ Bishop}

\author{M.\ Shayegan}

\affiliation{Department of Electrical Engineering, Princeton
University, Princeton, NJ 08544}

\begin{abstract}

We report effective mass (\textit{$m^{*}$}) measurements, via
analyzing the temperature dependence of the Shubnikov-de Haas
oscillations, for dilute, interacting, two-dimensional electron
systems (2DESs) occupying a single conduction-band valley in AlAs
quantum wells. When the 2DES is partially spin-polarized,
\textit{$m^{*}$} is larger than its band value, consistent with
previous results on various 2DESs. However, as we fully spin
polarize the 2DES by subjecting it to strong parallel magnetic
fields, \textit{$m^{*}$} is unexpectedly suppressed and falls even
below the band mass.

\end{abstract}


\maketitle

In a crystalline solid, electrons moving in the periodic potential
of ions are described as quasi-particles with a $band$ effective
mass, \textit{$m_{b}$}, which is inversely proportional to the
curvature of the energy-vs-wavevector (band) dispersion. In the
presence of electron-electron interaction, in the Fermi liquid
description, the electrons can still be treated as quasi-particles
but with a further re-normalized effective mass, \textit{$m^{*}$}.
Numerous studies, both experimental and theoretical, have indeed
reported that for low-disorder, dilute, two-dimensional electron
systems (2DESs), \textit{$m^{*}$} is typically larger than
\textit{$m_{b}$} at very low densities ($n$), and increases as $n$
is reduced and the system is made more interacting
\cite{smithPRL72,kwonPRB94,panPRB99,pudalovPRL02,shashkinPRB02,shashkinPRL03,vakiliPRL04,tanPRL05,zhangPRB05,asgariPRB05,gangadharaiahPRL05,zhangPRL05}.
Note that the parameter \textit{$r_{s}$}, defined as the ratio of
the Coulomb to kinetic (Fermi) energy, increases as the 2DES is made
more dilute \cite{footnote1}. Here we report \textit{m$^{*}$}
measurements in dilute 2DESs confined to AlAs quantum wells. Our
main finding, shown in Fig. 1, is that \textit{m$^{*}$} depends not
only on $n$ but also on the spin-polarization of the 2DES. When the
2DES is partially spin-polarized, the measured \textit{m$^{*}$} is
larger than \textit{$m_{b}$}, consistent with most previous reports
on other 2DESs. When we fully spin polarize the 2DES by applying a
parallel magnetic field, however, \textit{m$^{*}$} is strongly
suppressed to values below \textit{$m_{b}$}.


\begin{figure}
\includegraphics[scale=1]{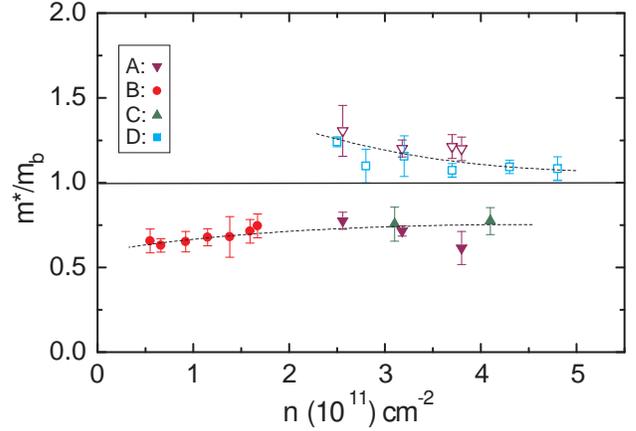}
\caption{ (color online). Measured effective mass, normalized to the
band mass, as a function of density for four different samples (A,
B, C, and D). The upper and lower branches represent masses measured
in partially and fully spin-polarized 2DESs, respectively. Each data
point represents \textit{m$^{*}$} averaged over different Landau
level filling factors ($\nu$), and the error bar includes the
variation of \textit{m$^{*}$} with $\nu$. The curves through the
data points are guides to the eye.}
\end{figure}

Our AlAs quantum well samples were grown, using molecular beam
epitaxy, on semi-insulating (001) GaAs substrates. The AlAs wells in
these samples have widths of 11 nm (sample A), 12 nm (sample B), or
15 nm (samples C and D). They are bounded by AlGaAs barriers and are
modulation-doped with Si \cite{shayeganPhysicaB06}. In our samples,
thanks to a combination of residual and applied uniaxial in-plane
strain \cite{shayeganPhysicaB06}, the electrons occupy one
conduction band minimum (valley) with an anisotropic (elliptical)
Fermi contour, characterized by transverse and longitudinal band
effective masses, \textit{m$_{t}$} = 0.205\textit{m$_{e}$} and
\textit{m$_{l}$} = 1.05\textit{m$_{e}$}, where \textit{m$_{e}$} is
the free electron mass. This means that the relevant
(density-of-states) band effective mass in our 2DES is
\textit{$m_{b}$} = $\sqrt{\textit{m$_{t}$}\textit{m$_{l}$}}$ =
0.46\textit{m$_{e}$}; this is the value to which we normalize and
report all our measured masses.

The samples were Hall bar or van der Pauw mesas, fitted with back
and front gates to tune \textit{n}. For the density range 0.55 - 4.8
$\times$ $10^{11}$ cm$^{-2}$, the low-temperature mobilities for the
samples are between 0.9 and 6 m$^{2}$/Vs when current is passed
along the low-mobility (longitudinal) direction of the occupied
valley; for current along the high-mobility (transverse) direction,
the mobility is typically 3 to 5 times larger than the above values.
Transport measurements were performed using standard low-frequency
lock-in techniques, and the samples were cooled in either a dilution
refrigerator or a $^{3}$He cryostat with base temperatures ($T$) of
20 mK and 0.3 K, respectively. In both cryostats, the samples were
mounted on a tilting stage so that the angle $\theta$ between the
sample normal and the direction of the magnetic field could be
varied; we define the total magnetic field as \textit{$B_{tot}$},
and the two components by \textit{$B_{\bot}$} and \textit{B$_{||}$}
(see Fig. 2 inset).

\begin{figure}[t] \centering
\includegraphics[scale=1]{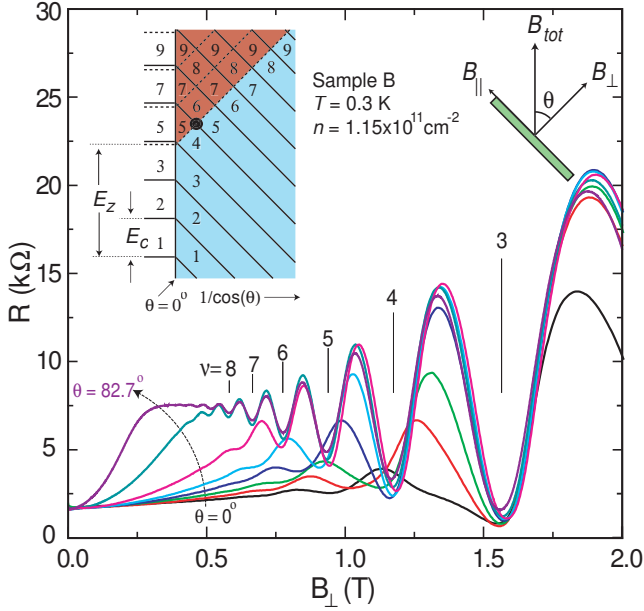} \caption{ (color online).\ Evolution of the
Shubnikov-de Haas oscillations with tilt angle, $\theta$, as
depicted by the top right inset. Magnetoresistance traces shown were
taken at $\theta$ = 0$^{\circ}$, 36.3$^{\circ}$, 44.3$^{\circ}$,
52.8$^{\circ}$, 59.6$^{\circ}$, 67.6$^{\circ}$, 77.0$^{\circ}$ and
82.7$^{\circ}$. Upper left inset shows the schematic fan diagram as
a function of tilt. The blue (light) and brown (dark) shaded regions
show the completely and partially spin-polarized regimes,
respectively. The angle beyond which the 2DES at $\nu$ = 5 becomes
completely spin-polarized is shown by a black circle.} \end{figure}

\begin{figure}[t] \includegraphics[scale=1]{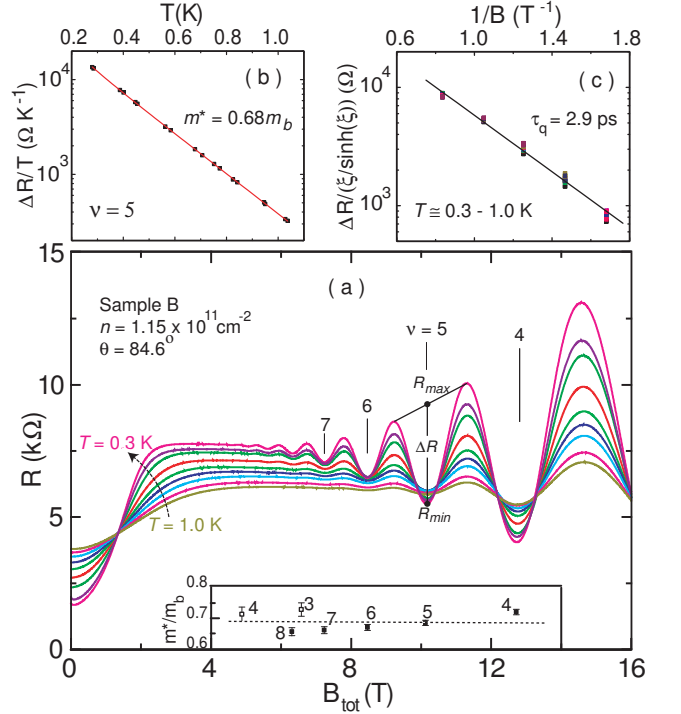}
\caption{ (color online).\ (a) $T$-dependence of SdH oscillations at
$\theta$ = 84.6$^{\circ}$. The traces were taken at $T\cong$ 0.30,
0.40, 0.45, 0.55, 0.70, 0.75, 0.80, 0.95, and 1.0 K. (b) Plot of
\textit{$\Delta$R/T} vs. \textit{T} and the fit (solid curve) to the
Dingle expression to extract \textit{m$^{*}$} for $\nu$ = 5. (c)
Dingle plot of $\Delta R/(\xi/sinh(\xi))$ vs. 1/$B$ summarizing data
taken in the range 0.3 $\lesssim T \lesssim $1.0 K and 4 $\leq \nu
\leq$ 8. Bottom inset to (a) shows the deduced \textit{m$^{*}$} as a
function of \textit{$B_{tot}$}. All data points are for the fully
spin-polarized 2DES. Closed squares represent \textit{m$^{*}$}
deduced from the $T$-dependence data shown in the main figure. Open
squares are from similar measurements at the same density at
$\theta$ = 75.3$^{\circ}$. For each data point, its $\nu$ is
indicated near it, and the error bar comes from the fit to the
Dingle expression, an example of which is shown in (b).}
\end{figure}

Figure 2 shows the evolution of magnetoresistance traces for sample
B at \textit{n} = 1.15 $\times 10^{11}$cm$^{-2}$ as it is tilted in
the magnetic field. The simple Landau level (LL) diagram shown in
the inset depicts the quantized energy levels split by the cyclotron
(\textit{E$_{C}$}) and Zeeman (\textit{E$_{Z}$}) energies and
explains this evolution. For this density, at $\theta$ =
0$^{\circ}$, \textit{E$_{Z}$} $\cong$ 3\textit{E$_{C}$}. This leads
to a LL "coincidence" \cite{fangPRL68} near $\nu$ = 4 and the
resistance exhibits a maximum rather than a minimum. As the sample
is tilted away from $\theta$ = 0$^{\circ}$, \textit{B}$_{||}$
increases and enhances the ratio of \textit{E$_{Z}$/E$_{C}$}, making
LLs of opposite spin go through coincidences. This causes, e.g., the
$\nu$ = 5 resistance minimum at $\theta$ = 0$^{\circ}$ to become a
maximum (at $\theta$ = 46.8$^{\circ}$) and then turn into a strong
minimum at larger $\theta$ when the 2DES becomes fully
spin-polarized. Note that for each $\nu$, there is a $\theta$ beyond
which the system becomes completely spin-polarized.

Data of Fig. 2 reveal a stark contrast between the traces taken at
small and large $\theta$: the resistance oscillation centered around
a given $\nu$ becomes much stronger once $\theta$ is increased
beyond the last coincidence angle (for that $\nu$) and the 2DES is
fully spin-polarized. This observation provides a qualitative hint
that the effective mass is suppressed when the spins are all
polarized.

To verify this conjecture we performed \textit{m$^{*}$} measurements
on this sample at the same density at a large tilt angle ($\theta$ =
84.6$^{\circ}$) as shown in Fig. 3(a) \cite{footnote2}. We followed
the usual procedure for mass determination from the
\textit{T}-dependence of the amplitude ($\Delta R$) of the
Shubnikov-de Haas (SdH) resistance oscillations. In the simplest
picture $\Delta R$ is given by the Dingle expression
\cite{dinglePRSL52}: $\Delta R/R_{o} = 8
exp(-\pi/\omega_{c}\tau_{q})\xi/sinh(\xi)$, where $R_{o}$ is the
non-oscillatory component of the resistance and $\tau_{q}$ is the
single-particle (quantum) lifetime. In the simplest picture, both
$R_{o}$ and $\tau_{q}$ are assumed to be $T$-independent. Here we
make this assumption and will return to its consequences later in
the paper. The factor $\xi/sinh(\xi)$ represents the \textit{T}
induced damping where $\xi=2\pi^{2}k_{B}T/\hbar\omega_{c}$ and
\textit{$\omega_{c}=eB_{\bot}/m^{*}$} is the cyclotron frequency. We
define $\Delta$\textit{R} = (\textit{$R_{max} - R_{min}$}) as
indicated in Fig. 3 (for the lowest $T$ trace), and plot
\textit{$\Delta$R/T} vs. \textit{T} on a semi-log plot and fit the
data to the above Dingle expression to determine \textit{m$^{*}$}.
Figure 3(b) shows such a plot for $\nu$ = 5. The values of
\textit{m$^{*}$} so obtained are shown in the inset to Fig. 3(a).
Evidently, \textit{m$^{*}$} for the fully spin-polarized 2DES is
well below the band mass ! In Fig. 3(c), in a "Dingle plot" of
$\Delta R/(\xi/sinh(\xi))$ vs. 1/$B$, we observe an approximately
linear behavior, implying a field-independent $\tau_{q}$ = 2.9 ps.

\begin{figure}[b]\includegraphics[scale=1]{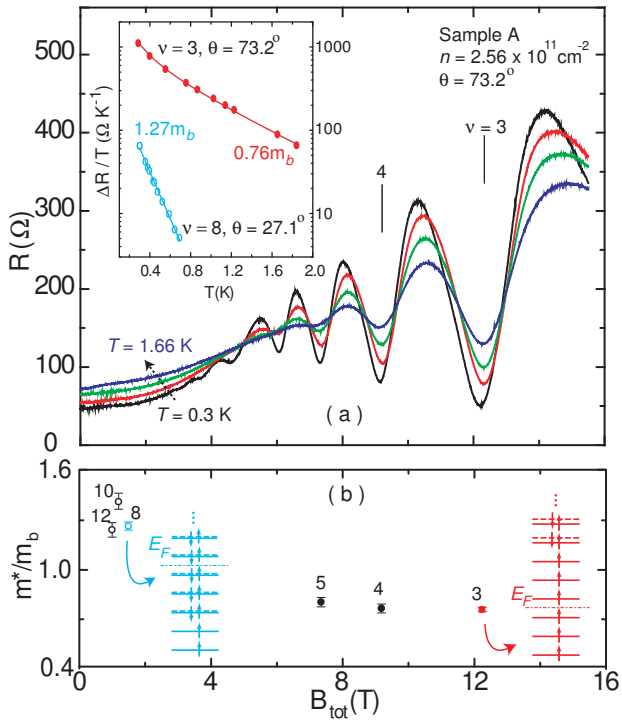}
\caption{ (color online).\ Dependence of \textit{$m^{*}$} on
spin-polarization for sample A at a fixed density of $n = 2.56
\times 10^{11}$ cm$^{-2}$. (a) $T$-dependence of SdH oscillations at
$\theta$ = 73.2$^{\circ}$. Traces were taken at $T\cong$  0.30,
0.87, 1.23, and 1.66 K. Inset: Plots of (\textit{$\Delta R/T$}) vs.
\textit{T} and fits to the Dingle expression at $\nu$ = 3 (fully
spin-polarized) and $\nu$ = 8 (partially spin-polarized); note that
the $\nu$ = 8 data were taken at $\theta$ = 27.1$^{\circ}$. In (b),
we show the deduced \textit{$m^{*}$} as a function of
\textit{$B_{tot}$}. The closed and open circles correspond to the
full and partial spin-polarizations, respectively. The two ladder
diagrams depict the positions of the Landau levels and the Fermi
energy (\textit{$E_{F}$}) for the cases of $\nu$ = 3 and 8 whose
data are shown in the inset to (a).}
\end{figure}

Next, we present data showing the variation of \textit{m$^{*}$} with
spin-polarization at a fixed $n$. In Fig. 4(a) we show traces for
sample A taken at $n = 2.56\times 10^{11}$cm$^{-2}$ and $\theta$ =
73.2$^{\circ}$. At this $\theta$, the 2DES is fully spin-polarized
for $\nu < 7$. We show the measured \textit{m$^{*}$} for $\nu$ = 3,
4, and 5 in Fig. 4(b); the inset to Fig. 4(a) shows the
$T$-dependence of \textit{$\Delta$R/T} for $\nu=3$. In Fig. 4(b) we
also include \textit{m$^{*}$} deduced for the partially
spin-polarized case from similar magnetoresistance data (not shown)
at $\theta$ = 27.1$^{\circ}$; the \textit{$\Delta$R/T} vs.
\textit{T} plot for $\nu=8$ for this $\theta$ is also shown in Fig.
4(a) inset. Clearly, \textit{m$^{*}$} in the partially polarized
case is much larger than \textit{m$^{*}$} for the fully polarized
case. The energy ladder diagrams in Fig. 4(b) schematically show the
positions of the various LLs and the Fermi energy ($E_{F}$) for two
representative cases whose data are shown in Fig. 4(a) inset. Note
that when the 2DES is partially spin-polarized, its LLs could be
staggered so that \textit{$E_{F}$} can be in an energy gap whose
magnitude is smaller than $\hbar\omega_{c}$. In our experiments, for
each partially spin-polarized case, we carefully chose $\theta$ (via
coincidence measurements) to ensure that the spin-up and spin-down
levels overlap, as shown in the left ladder in Fig. 4(b). Note also
that for the case shown ($\nu=8$), the 2DES spin-polarization $P$,
defined as the difference between the number of occupied up and down
spin levels divided by the total number of occupied levels, is $P =$
0.25.

Figure 1 summarizes our results for four samples of different
well-widths, densities, and mobilities, measured in seven separate
cooldowns, and over a wide range of $\nu$. The data also include
different measurement geometries, e.g., orienting the current and/or
\textit{B$_{||}$} along the major or minor axis of the occupied
conduction band Fermi ellipse. For example, for sample A,
\textit{B$_{||}$} was applied along the minor axis of the occupied
conduction-band ellipse while for B, C and D, \textit{B$_{||}$} was
along the major axis. As is clear from Fig. 1, the \textit{m$^{*}$}
suppression is observed in $all$ cases and is independent of such
parameters. We have preliminary data indicating that the
\textit{m$^{*}$} suppression is also observed in fully
spin-polarized $narrow$ AlAs quantum wells (well width $<$ 5nm)
where the 2D electrons occupy an out-of-plane conduction band valley
with an $isotropic$ in-plane Fermi contour and \textit{$m_{b}$}=
0.205\textit{m$_{e}$}. The fact that we observe the suppression in
both of these two 2DESs suggests that the suppression is general.

However, a few words of caution regarding \textit{m$^{*}$}
determination from the amplitude of the SdH oscillations are in
order. As is generally done, in our analysis we have assumed that
\textit{$R_{o}$} and $\tau_{q}$ or, equivalently, the Dingle
temperature $T_{D} = \hbar/2 \pi k_{B} \tau_{q}$, are
$T$-independent. Such an assumption indeed appears reasonable for
data of Fig. 4(a) where the background resistance in the field range
where we analyze the $T$-dependence of the SdH oscillations is
essentially independent of $T$. But note in Fig. 3 that the
background resistance depends on $T$. More generally, we typically
observe an enhancement of \textit{m$^{*}$} (over \textit{$m_{b}$})
when the $T$-dependence of the background resistance is "metallic",
while the \textit{m$^{*}$} suppression ensues when the 2DES has
turned "insulating" following the application of a sufficiently
large \textit{B$_{||}$} to fully spin-polarize it (e.g., in Fig. 3,
the 2DES is fully spin-polarized for $B_{tot} > 2.3$ T). To check
the possible consequences of the $T$-dependence of resistance
background on the deduced \textit{m$^{*}$}, we analyzed our data in
two additional ways. Suppose we assume that $T_D$ is independent of
$T$ but, to account for the $T$-dependence of the background
resistance, we define \textit{$R_{o}$} as the average resistance
near the oscillation, i.e., \textit{$R_{ave}$} = (\textit{$R_{max}$}
+ \textit{$R_{min}$})/2 (see Fig. 3(a)). Such a procedure leads to
\textit{m$^{*}$} values which are about 6$\%$ smaller than those
indicated in Fig. 3. In a third scenario, we might assume that
$R_{ave}$ and $T_{D}$ have the same $T$-dependence; this may be a
reasonable assumption, since $R_{ave}$ and $T_{D}$ are both expected
to be proportional to the scattering rate. To implement such an
assumption, we use an iterative procedure to find a value for
\textit{m$^{*}$} that leads to a $T$-dependence for $T_D$ such that
the ratio $R_{ave}$/$T_{D}$ is $T$-independent. We find that
\textit{m$^{*}$} from this method are about 11$\%$ larger than
\textit{m$^{*}$} indicated in Fig. 3. We emphasize that $all$ the
\textit{m$^{*}$} values reported in our manuscript come from data
sets whose analysis via the three procedures described above yield
consistent \textit{m$^{*}$} values. In particular, our main
conclusion that \textit{m$^{*}$} for the fully spin-polarized case
is suppressed is borne out in all three procedures.

Interaction-induced re-normalization of \textit{m$^{*}$} in dilute,
interacting 2DESs has been studied widely
\cite{smithPRL72,kwonPRB94,panPRB99,pudalovPRL02,shashkinPRB02,shashkinPRL03,vakiliPRL04,tanPRL05,zhangPRB05,asgariPRB05,gangadharaiahPRL05,zhangPRL05,coleridgeSS96}.
Experimental studies on partially spin-polarized 2DESs confined to
Si-MOSFETs
\cite{smithPRL72,panPRB99,pudalovPRL02,shashkinPRB02,shashkinPRL03}
or narrow AlAs quantum wells \cite{vakiliPRL04} have reported that
\textit{m$^{*}$} is significantly enhanced with respect to $m_{b}$
and increases as $n$ is lowered. For GaAs 2DESs, too, an enhancement
of \textit{m$^{*}$} has been reported at very low densities ($r_s
\gtrsim$ 3) \cite{tanPRL05}, but, at high densities \textit{m$^{*}$}
is slightly suppressed compared to $m_{b}$
\cite{tanPRL05,coleridgeSS96}; there is also some qualitative
theoretical explanation for this non-monotonic behavior (see Ref.
\cite{tanPRL05}).

It is intuitively clear that, besides $r_s$, the spin-polarization
of the 2DES should also affect the \textit{m$^{*}$} re-normalization
since it modifies the exchange interaction. However, most previous
studies have ignored this role. One notable exception is the
experimental work by Shashkin \textit{et al.} \cite{shashkinPRL03}
where measurements of \textit{m$^{*}$} revealed no dependence on
spin-polarization for $0 < P \leq 1$. In our study, the
"partially-spin-polarized" regime corresponds to $0.14 < P < 0.33$
and indeed, in this range, we find no dependence of \textit{m$^{*}$}
on $P$. We observe the strong mass suppression only when the 2DES is
pushed well beyond the full spin-polarization field limit, i.e. when
the applied magnetic field is well above what is needed for complete
spin-polarization \cite{footnote3}. For example, for the densities
in Figs. 3 and 4, the fields needed for complete spin-polarization
of the 2DES are around 2.3 T (Fig. 3) and 5.7 T (Fig. 4), whereas
our first suppressed mass is reported at around 4.9 T and 7.3 T,
respectively.

Motivated by the results of Ref. \cite{shashkinPRL03}, recent
theoretical work \cite{gangadharaiahPRL05,zhangPRL05} has addressed
the role of spin-polarization on \textit{m$^{*}$} re-normalization.
In Ref. \cite{gangadharaiahPRL05} it is concluded that, although
\textit{m$^{*}$} should in principle depend on spin-polarization, in
a valley-degenerate system such as the one studied in Ref.
\cite{shashkinPRL03}, the dependence may be too weak to be
experimentally measurable \cite{footnote4}. In Ref.
\cite{zhangPRL05}, which deals with a single-valley 2DES, however, a
rather strong dependence of \textit{m$^{*}$} on spin-polarization is
reported. There is even a hint that for a fully spin-polarized 2DES
\textit{m$^{*}$} may fall below \textit{$m_{b}$}. But there are some
qualitative discrepancies between the predictions of Ref.
\cite{zhangPRL05} and our experimental results. For example,
\textit{m$^{*}$} for a fully spin-polarized 2DES is predicted to
decrease with $increasing$ $n$ and become smaller than
\textit{$m_{b}$} only at $large$ values of $n$ (\textit{$r_{s}$}
$\lesssim$ 2) \cite{zhangPRL05}. In contrast, our data suggest that
the \textit{m$^{*}$} suppression is observed in a large density
range and may even be more pronounced at $smaller$ densities. An
understanding of the \textit{m$^{*}$} suppression we observe in a
dilute, single-valley, fully spin-polarized 2DES therefore awaits
future theoretical and experimental developments.

We thank the NSF for financial support. Part of this work was done
at the NHMFL, Tallahassee, which is also supported by the NSF. We
thank E. Palm, T. Murphy, J. Jaroszynski, S. Hannahs and G. Jones
for assistance. We also express gratitude to A.H. MacDonald, E.
Tutuc, and R. Winkler for illuminating discussions.

\end{document}